# Fluid dynamics: an emerging route for the scalable production of graphene in the last five years

Min Yi[1] and Zhigang Shen[2]

[1] Institute of Materials Science, Technische Universität Darmstadt, Darmstadt 64287, Germany.

[2] Beijing Key Laboratory for Powder Technology Research & Development, Beihang University (BUAA), Beijing 100191, China.

E-mail: yi@mfm.tu-darmstadt.de; shenzhg@buaa.edu.cn

## Abstract

Bulk applications of graphene in fields such as advanced composites, conductive ink, and energy storage require cheap and scalable graphene. Fortunately, in the last decade, liquid-phase exfoliation of graphite to give pristine graphene has been thought as a promising way to massive production of graphene at high efficiency and low cost, in terms of the cheap and abundant graphite source and a variety of cost-effective exfoliation techniques. Though many exfoliation techniques are available so far, this article will highlight the recent progress of fluid dynamics route which emerges as a promising scalable and efficient way for graphene production in the last five years. The emphasis is set on vortex fluidic devices and pressure- and mixer-driven fluid dynamics, with our perspectives on the latest progress, exfoliation mechanism, and some key issues that require further study in order to realize industrial applications.

## Contents

# 1. Introduction

Due to its exceptional properties and intriguing applications, graphene has attracted intensive interests in the advanced science and technology. The last decade has witnessed many breakthroughs in research on graphene. A serial of novel properties and promising applications are reported in succession [1-7]. With the achievement by the graphene community in recent years, it has been thought that graphene will become the next subversive technology, substitute some of the currently used materials, and result in a new market and thus the scientific and technological revolution. However, the scalable and cost-effective production of graphene still remains as a critical issue for realizing its commercialization. If graphene cannot be produced at low cost and high efficiency, its commercial and widespread use could be lowered down or even ultimately hindered.

Since graphene was discovered in 2004 [8], a significant advance in the mass production of this material has been achieved, as shown in **Fig. 1**. A great many methods have been proposed to produce graphene [9-35], among which one can choose a suitable one for specified applications. The bottom-up methods [10, 15, 18, 23, 25-35] such as chemical vapour deposition and epitaxial growth can produce high quality, large-size and thickness-controllable graphene. The resulted graphene is ideal for fabricating graphene electronics, field-effect transistors, flexible transparent electrodes, functional touch-screen panel device, etc. However, these substrate-based techniques suffer from the limited scale and high cost, and cannot meet the requirement of macroscopic quantities of graphene for applications such as advanced composites, coatings, conductive ink, and energy storage. Fortunately, liquid-phase exfoliation (LPE) of graphite to give graphene has been recently proved as a scalable method [2, 3, 9, 12, 13, 16, 17, 19-22, 36-87]. This method uses the cheap and abundant graphite flakes as the precursor. The graphene products generated by this method can fit the requirement of scalability, reproducibility, processability, and low production cost. Though the graphene quality by this method is a question, the definition on the graphene quality should be highly dependent on the specified application. For example, in the applications for high catalytic activity [88, 89] and high storage of capacitive charges [90], the graphene with edge defects are preferable. Graphene nanomesh with porous structure is desired for semiconductivity applications which require a tunable bandgap [78, 91-93]. Therefore, the LPE method is very promising. It should be noted that the LPE method depends on the exfoliation medium and the exfoliation technique. The exfoliation medium, such as suitable organic solvents, surfactant/water solutions, aromatic solvents, ionic liquid, etc., has been discussed in several art-to-date review [9, 12, 14, 16, 17, 20]. The widely used exfoliation technique is sonication, which has also been reviewed recently [13, 41, 94]. The sonication depends on the liquid cavitation for exfoliation. However, sonication induced cavitation is a relatively harsh process which can produce high local temperature (~ several thousand K), extreme pressure (~ several thousand atm), and rapid heating/cooling rates (~ several billion K/s) [95-97]. These harsh conditions involved in cavitation could result in damage to the graphene. Thus, the graphene produced by sonication has been verified to have much more defects as expected [74, 98-100]. Moreover, the distribution and intensity of the sonication-induced cavitation are highly dependent on the vessel size and shape which often induce localized cavitation pictures [101-104]. If the position of the ultrasonic vibration source is fixed, the cavitation field in the liquid is almost static. These drawbacks are not favorable for efficient exfoliation and a large quantity of graphite flakes which settle down to the bottom still remain unexfoliated. In addition, in the literatures, sonic tips can only effectively process volumes no larger than a few 100 mL leading to low production rates. While sonic baths can be used to process hundreds of millilitres, the power transfer from bath to liquid is relatively poor, leading to long exfoliation times and so low production rates. Another route for efficiently transferring mechanical energy directly into the liquid is desired.

Apart from sonication, fluid dynamics has emerged as a novel exfoliation technique for scalable production of graphene in the recent five years. Within the fluid dynamics, graphite flakes can move with the liquid and thus can be exfoliated repeatedly at different position. And multiple fluid dynamic events are responsible for exfoliation. These features are intrinsically different from that of sonication, rendering it as a potentially efficient technique. Hence, keeping those key factors in mind, this progess report will examine three promising fluid dynamics based exfoliation routes (vortex fluidics, pressure and mixer driven fluid dynamics) and highlighting their recent progress and challenges.

## 2. Vortex fluidic device (VFD)

In order to avoid the graphene defects and solvents degradation induced by cavitation in sonication, a VFD is developed to generate a less energy intensive shear process for exfoliating graphite. The VFD is schematized in **Fig. 2**a. It consists of a tube open at one end. When it is rapidly rotated, intense shear will be generated in the resulting thin films with finite sub-millilitre volumes of liquid. The shear fluidic film can be controlled by adjusting the speed and orientation of the tube, and other operating parameters [105]. According to the fluid dynamics [106], a rapidly rotating fluid can generate the boundary and shear layers parallel to the axis of rotation, named as Stewartson/Ekman layers. Within this layer, the liquid flow is upwards at the internal surface of the rotating tube, and downwards close to the liquid surface, as shown in **Fig. 2**a. Graphene dispersed in N-methyl-pyrrolidone (NMP) was successfully produced via exfoliating graphite by the shear vortex fluidic films in the 'confined mode' of operation of the VFD (without jet feeds in **Fig. 2**a).[58]. The graphite flakes dispersed in NMP will initially accelerate to the walls of the tube by the large centrifugal force. Then the partial lifting and slippage on the tube wall are responsible for the exfoliation mechanism, as shown in **Fig. 2**b and c. The slippage process can be highlighted by the "finger print" of partially stacked graphene in **Fig. 2**e. This slippage process requires the individual sheets to be partially lifted from the surface of the bulk material at some point to provide the necessary lateral force to start the slippage (**Fig. 2**b). Meanwhile, the graphite flakes were pushed against the tube wall by the centrifugal force and experienced a shear induced displacement along the tube, resulting in exfoliation at the tube surface (**Fig. 2**c).

In contrast, by using the 'continuous flow mode' of the VFD, graphene based hybrid materials [72] and functionalized graphene [107] can also be readily produced. In the 'continuous flow mode', another jet feeds can deliver liquid into the rotating tube (**Fig. 2**a). This will generate additional shear is in the thin films by the viscous drag as the liquid whirls along the tube. As shown in **Fig. 3**, the 'confined mode' is firstly used to exfoliate graphite into multi-layer graphene in water. Then in the 'continuous flow mode', a feed jet at the base of the tube is used to deliver the recirculating liquid of graphene and microalgae mixed suspension. With this route, the multi-layer graphene sheets can be decorated on the surface of microalgal cells.

Following the same idea, Tran et al. [108] applied a Taylor-Couette flow reactor to generate the vortex flow for exfoliating graphite flakes, as schematized in **Fig. 4**. In the reactor, the mixture of graphite and solvent is sheared between a rapidly rotating inner cylinder and stationary outer cylinder. Thus the vortex flow induced high wall shear stress and pressure can be utilized to exfoliate graphite flakes. It is demonstrated that this method can efficiently produce few-layer graphene with low degree of defects. However, the gap between these two cylinders is only 2.5 mm, limiting the throughput.

This VFD offers an alternative and tunable low-energy source for mild exfoliation and thus high-quality graphene. But the vortex fluidic film or the gap between is extremely thin (in the millimeter order), which limits the quantity of graphite used for exfoliation and the graphene output.

The lateral size of graphene prepared by VFD is generally less than 1 μm, as shown in **Fig. 2**e-g and **Fig. 5**a and b. This is due to that the weak shear force generated by VFD can only exfoliate smaller graphite flakes; because the collective van der Waals interaction between layers for larger graphene flakes is much higher. The thickness of graphene prepared by VFD changes from less than 1 nm to more than 20 nm, as shown in **Fig. 2**f and g and **Fig. 5**b. However, the number of these transmission electron microscopy (TEM) and atomic force microscopy (AFM) images in the literatures are too small to give a statistical analysis of the distribution of sheet size. Since the graphene lateral size and thickness can determine whether graphene can be integrated into practical devices and its ultimate properties are attainable, it is highly recommended to obtain the size distribution of graphene prepared by VFD in the near future. As for the defects, it is anticipated that VFD generates weak shear force and it will not damage graphene. But there are only Raman spectra results (**Fig. 5**c) in this aspect. A deeper study by means of different microscopic and spectral techniques is required for the determination of defects or oxides in the graphene prepared by VFD, in order to establish VFD as a really defect-free method.

## 3. Pressure-driven fluid dynamics (PFD)

In order to realize large-scale production of graphene, the pressure-driven fluid dynamics (PFD) is utilized [55, 56, 61, 69, 75, 78, 84, 86, 109]. Unlike the VFD which depends on the rapidly rotating tube, PFD relies on a serial of flow channels for exfoliation. The pressure difference between the inlet and the outlet can generate rich fluid dynamics events in the flow channels. Within the PFD device, graphite flakes can move with the liquid along the flow channel and thus can be exfoliated repeatedly at different position. This is totally different from the case in sonication where the location of the cavitation field and exfoliation events is almost static. This feature renders PFD as a much more efficient technique. The flow channel can be in the order of either micrometer [75, 84, 109] (**Fig. 6**a and **Fig. 7**) or millimeter [55, 56, 61, 69, 78, 86] (**Fig. 6**b). The number of the expansion and contraction channels in PFD devices can be adjusted during the manufacturing. The fluid dynamics in the PFD is featured by cavitation, pressure release, viscous shear stress, turbulence, and collision. As illustrated in **Fig. 6**c, there are multiple fluid dynamics events responsible for normal- and shear-force dominated exfoliation. Cavitation and pressure release can generate normal force for exfoliation. The velocity gradient-induced viscous shear stress, the turbulence-induced Reynolds shear stress, and shear effects stemmed from turbulence and flow channel-induced collisions can generate shear force for exfoliation, resulting in graphite self-exfoliation down to single or few layers through its lateral self-lubricating ability. All these dynamics events have fragmentation effects which also facilitate exfoliation; because the collective interaction force between two adjacent layers is smaller in the smaller flakes. Hence, the PFD possesses multiple ways for exfoliation, providing great advantages over sonication which possess a single cavitation effect and ball milling or VFD which possesses single shear effect, in terms of yield and efficiency.

The produced graphene can be verified by different characterizations, such as scanning electron microscopy (SEM) and AFM (**Fig. 6**d-f). By capturing a large number of AFM images, the lateral size (or flake area) and thickness distribution of graphene prepared by PFD can be obtained. Moreover, by adjusting the pressure and treating time, the graphene concentration, thickness distribution and flake area distribution can be controlled, as shown in **Fig. 8** [86]. Depending on the pressure and treating time, the flake area and thickness distribution can be changed. As the treating time is increased, the thickness distribution becomes narrower and shifts to lower values **Fig. 8**a). For example, in the 0.5 h sample, flakes with thickness <3 nm (less than 10 layers) occupies ~80%. The percentage of thin flakes with thickness <1.5 nm (less than 5 layers) of the 0.5 h, 4 h, and 8 h samples is 29%, 63%, and 79%, respectively. In contrast, the flake area sharply moves to small values. **Fig. 8**b presents that over 85% flakes are with area less than $10^5$ nm$^2$ in the 8 h sample. The mean value of the flake area has decreased by an order of magnitude in comparison with that in the 0.5 h sample. These results on size distribution establish PFD as a controllable method for preparing graphene with specified size. As for the defects, only Raman results on the graphene-based films are available [86, 109], which indicate low-level basal plane defects. Nevertheless, microscopic study on the individual flakes by scanning tunneling microscopy (STM) and X-ray photoelectron spectroscopy (XPS) is still required to get more detailed information on the atomic structure and chemical components of the basal plane of graphene prepared by PFD.

**Fig. 8**c shows that higher pressure leads to higher yield of graphene. These results are useful for scaling up this technique from 10 L per pot in laboratory to several hundred liters in industry. Most interestingly, if pressure is increased to higher values (e.g. 30 MPa), this PFD technique can be used to produce graphene nanomesh [78], which recently emerges as a novel graphene nanostructure with bandgap that is large enough for room-temperature transistor operation [91, 92]. The mechanism is the combination of exfoliation and perforation of the graphene sheets (**Fig. 9**a). The obtained graphene nanomesh is shown in **Fig. 9**b and c. It is estimated that the total area of the pores within 1 μm$^2$ nanomesh is ~0.15 μm$^2$ and the pore density is ~22 μm$^{-2}$. This provides a novel route for large-scale production of graphene nanomesh.

## 4. Mixer-driven fluid dynamics (MFD)

Another recently emerging method is the mixer-driven fluid dynamics (MFD). The device for realizing this method is relatively simple and easily available. A commercial available rotor/stator mixer can be used for graphene production [79, 80], as shown in **Fig. 10**. The head of the mixer is constituted by a rotor and a stator as the critical component for exfoliation. The rotor diameters (**Fig. 10**b and c) can be adjusted. By using NMP as the solvents, the mixer can result in graphene-NMP dispersions (**Fig. 10**d), in which graphene flakes are with lateral size of several hundred nanometers, as shown the TEM image in **Fig. 10**e and f. The shear exfoliation mechanism can be further revealed in terms of the rotor diameter and the mixer-induced fluid dynamics. It was found that even when the Reynolds number $Re_{Mix}$ of the flow field is less than $10^4$, which corresponds to a not fully developed turbulent flow, well-exfoliated graphene still can be obtained, as shown the region below the $Re_{Mix}$ line in **Fig. 10**g. But when the shear rate $\dot{\gamma}$ is lower than $10^4$ s$^{-1}$, graphite flakes are poorly exfoliated, as shown the region below the $\dot{\gamma}=10^4$ s$^{-1}$ line in **Fig. 10**g. In the case of the mixer at a number of different combinations of rotating speed and rotator diameter, the minimum shear rate $\dot{\gamma}_{min}$ is also around $10^4$ s$^{-1}$ (**Fig. 10**h). This suggests that any mixer that can achieve the shear rate above $10^4$ s$^{-1}$ can be used to produce graphene. The exfoliation mechanism can also be qualitatively explained in terms of the fluid dynamics events [79], as illustrated in **Fig. 11**. Like ball milling and VFD, this is a shear-force dominated method. But the cavitation and collision effects also favor efficient exfoliation (**Fig. 11**). However, in the rotor-stator mixer (**Fig. 10**b and c), very high shear rates are mainly localized in the gap between the rotor and stator and in the holes in the stator. This implies a well-defined localized region of high shear rate, indicating that most of the exfoliation events are localized in the vicinity of the rotor-stator.

Seeing that the high shear rates are mainly localized around the rotor/stator structure, a mixer equipped with rotating blades are proposed to induce fully developed turbulence to generate high shear rate all over the flow field. The simplest way to realize this is using a kitchen blender [71, 81, 83, 110], as shown in **Fig. 12**a and **13**a. In the kitchen blender, if the rotating speed of the blades is sufficiently high, the high-shear region is not localized in any single portion of the holder. Though the shear rate decreases with the increasing distance from the blade, high shear rate can cover all the region of the holder if a turbulence is fully developed. Therefore, the turbulence is mainly responsible for the full-field high shear rate and thus the exfoliation mechanism, as shown in **Fig. 12**b. In terms of the characteristics of the turbulent flow in the kitchen blender, it is demonstrated that four fluid dynamics events responsible for the exfoliation and fragmentation: (I) velocity gradient can induce viscous shear stress; (II) intensive velocity fluctuations in turbulence can induce Reynolds shear stress; (III) in the turbulence, Reynolds number is very large, and thus the inertial forces dominate viscous forces to enhance graphite-graphite collisions; (IV) it is possible that turbulent pressure fluctuations induced pressure difference can also exfoliate graphite in a normal-force style. The mechanism can be verified by the TEM observations. The slipped configuration with lateral relative displacement of translation (**Fig. 12**c) or rotation (**Fig. 12**d) indicates that lateral exfoliation really happens and there coexist two ways, i.e. translation and rotation. The exfoliation efficiency is much higher than that in standard sonication or ball milling exfoliation methods.

By using the kitchen blender, even the household detergent can be the surfactant for graphene production [81], as shown the presence of significant amounts of foam in **Fig. 13**a. The impeller is equipped with four blades (**Fig. 13**b). Most of the as-produced graphene flakes are folded, as shown the TEM image in **Fig. 13**c. This is different from the case of sonication in which graphene flakes are sometimes folded. The larger fraction of folded flakes in mixer-exfoliated graphene relative to sonication exfoliated graphene reflects differences in the fluid dynamics of the two systems. By monitoring the graphene concentration under different blade rotating speed, a critical blade speed of around 2 krpm from this special kitchen blender in **Fig. 13**a can be determined, as shown in **Fig. 13**d. This knowledge is important for designing the rotating-blade mixer for large scale graphene production. The kitchen blender and the household detergent make the MFD route extremely simple and cost-effective.

For the large scale production of graphene for biological applications, Pattammattel and Kumar [110] applied kitchen blender to exfoliate graphite in protein solutions, as schematized in **Fig. 14**. Dependent on the charge of the protein used, the exfoliation efficiency is violently varied. Among the five proteins: BSA (bovine serum), β-lactoglobulin (bovine milk), lysozyme (egg white), ovalbumin (egg white), and hemoglobin (bovine blood), the strongly negatively charged BSA gives the highest efficiency. By using the BSA aqueous solutions, the kitchen blender can achieve exfoliation efficiency more than 4 mg/mL/h and a maximum concentration of 7 mg/mL. The BSA-coated graphene with controllable surface charge is shown to be stable under PH values of 3-11 and temperatures of 5-50 $^0$C. The combination of kitchen blender and proteins makes MFD as an effective tool for scalable and biological production of graphene in water.

As for the size distribution of graphene flakes prepared by MFD, Varrla et al. [81] and Yi et al. [83] have used AFM to perform statistical analysis. Varrla et al. [81] adopted a rotation speed of 18 krpm and a treatment of 60 min to prepare graphene. The AFM-based statistical distributions of length (L) and layer number (N) are shown in **Fig. 13**e and f, respectively. The average length is found to be around 320 nm. The average layer number is around 6 [81]. In contrast, Yi et al. adopted a rotation speed of 5 krpm, and investigated the effect of preparation time on the size distribution of the resultant graphene flakes [83]. **Fig. 15** shows the AFM-based statistical results for flakes' dimensions. Area rather than length or width is chosen, because most graphene flakes are irregularly shaped and measuring their length or width is difficult. As shown in **Fig. 15**a, the number fraction of ≤1.5-nm-thick flakes exceeds 80% for all the preparation time, reaching a high value of ~92% at 3 h. Additionally, the number fraction of ≤1 nm-thick graphene flakes approximately keeps constant between 14.6% and 20% for all the preparation time. The flake area notably decreases with preparation time, resulting in a shift in the area distribution towards lower values, as shown in **Fig. 15**b. Based on these statistical data, the average thickness per flake, <t>, and the average area per flake, <A>, can be calculated, as shown in **Fig. 15**c. <t> hardly varies at all and maintains at ~1.5 nm, corresponding to an average layer number of <5. Nevertheless, the thickness distribution shifts towards lower values as preparation time increases, as illustrated in **Fig. 15**a. In contrast, <A> decreases with preparation time, falling from ~2.4 $\mu m^2$ at 0.5 h to ~0.1 $\mu m^2$ at 8 h. By fitting <A> as a function of preparation time, an inversely-proportional relationship appears, as shown in **Fig. 15**d. For the biographene produced in water/proteins solutions by MFD in the kitchen blender, Pattammattel and Kumar [110] calculated the average layer number and lateral size as a function of exfoliation time, as shown in **Fig. 16**b and c. The average layer number is estimated as ~3.6 in despite of the exfoliation time (**Fig. 16**b). The average size is ~0.5 μm and appears to be highly uniform, different from the results of Varrla et al. [81] and Yi et al. [83]. The reason is attributed to the method. Pattammattel and Kumar [110] used the Raman data to obtain the size information by an empirical equation. In contrast, Varrla et al. [81] and Yi et al. [83] directly measured the size by AFM or TEM.

The defects and oxides of graphene prepared by MFD were studied by XPS and Raman mapping, as shown in **Fig. 16** [110] and **Fig. 17** [83]. By statistical analysis of the Raman intensity ratio, Pattammattel and Kumar [110] pointed that the biographene produce by MFD is only of minor edge defects (**Fig. 16**d). The XPS results in **Fig. 17**b show the same bonds and similar composition in the pristine graphite and graphene-based film, indicating that the low level of oxides in graphene are caused not by residual solvent or oxidation but by water, $CO_2$ or oxygen from the atmosphere. These prove that MFD does not chemically functionalize the graphene flakes. By using Raman mapping technique, individual graphene flake can be captured and its Raman spectrum can be obtained, as shown in **Fig. 17**a and c. In **Fig. 17**c, the 2D bands in Raman spectra of flake #1 and #2 reflect the graphene nature [22, 111, 112]. However, there are no D bands and the G bands are not remarkably widened. This indicates that the flakes are almost free of basal plane defects. For the precise examination of the local defects or atomic structure, STM and high-resolution TEM characterizations are further required.

## 5. Conclusions and perspectives

Since the second half of 2011 when the virgin idea of producing graphene by utilizing the rich flow events in fluid dynamics was firstly initiated,[55, 56] huge progress has been made during the last five years. Various methods for generating fluid dynamics have been proposed in order to explore an efficient and scalable route for graphene production, such as vortex fluidics, mixer, blender, high pressure, etc. Compared to the widely used cavitation-dominated sonication for producing graphene, fluid dynamics possess multiple exfoliation effects originated from the shear, cavitation, collision, and pressure release. Therefore, fluid dynamics are far more efficient than sonication routes and shows great technological potential in the near future. Considering the main factors for industrialization of graphene production, i.e. production efficiency, production cost, scalability, reproducibility, processability, etc., the recently emerging fluid dynamics route is very promising. With the continuous effort in fluid dynamics for graphene production, many exciting results and new methods have so far been reported and several technologies are currently envisioned. We believe that the recently emerging fluid dynamics route provides a significant step in the direction of making the commercial availability of large quantities of high-quality graphene. To proceed from discovery to a commercialized technology, many issues remain to be further explored, a partial list of which includes:

(1) How can we control and optimize the exfoliation effects in fluid dynamics, so that the harsh and violent effects can be lowered to a minimum level? For example, though cavitation can exfoliate graphite into graphene flakes, it induces extremely high local temperature and pressure [95-97], which can result in defected graphene. In contrast, relying on the lateral exfoliation mechanism, exfoliation by shear force is much milder. In PFD and the local region near the rotating blade in MFD, high-speed fluid can generate cavitation. A deep understanding and precise design of the flow field in PFD and MFD are critical for eliminating the cavitation region and achieving high shear rates throughout the flow field.

(2) How to achieve monolayer dominated and large-size graphene products by fluid dynamics still remains challenging. Exfoliation in fluid dynamics is always accompanied with fragmentation that is not desired for producing large-size graphene. How to minimize the fragmentation effects should be considered. The average layer number and lateral size of graphene produced by fluid dynamics are 3-5 and several hundred nanometers, respectively. It indicates the case of few-layer graphene. The control of graphene size may be possible by combining fluid dynamic methods and specified centrifugation strategy [59, 113-116].

(3) The nature of defects induced by fluid dynamics requires detailed investigations. Currently a consensus has been reached on the conclusion of edge-dominated defects in graphene produced by fluid dynamics. But the conclusion is almost based on the Raman spectra of filtered graphene films, not the single graphene flake. In the filtered film, the Raman signal is a superposition of contributions from many single- and few-layer graphene flakes. It is suggested microscopic study on the individual flakes by STM and XPS to be carried out, in order to get more detailed information on the atomic structure and chemical components of the basal plane of graphene prepared by fluid dynamics.

(4) Other simple routes for generating fluid dynamics should be explored. As the schematics shown in **Fig. 18**, random shake and liquid spray may be another two possible routes. It is anticipated that the fluid dynamics events involved in the process of random shake and liquid spray can generate viscous shear, turbulence, collision, and pressure release to mildly exfoliate graphite into graphene flakes.

(5) Are the current fluid dynamic methods ready for industrial scale-up or 'blip on the oscilloscope'? The VFD route with weak shear force can prepare high-quality graphene, but the small throughput limits the scalable production. Though with high efficiency, the PFD device depends on high pressure and small flow channels, increasing the complexity and cost of the device. The MFD route is much simpler and the device for MFD is much more easily available. As the rotating blade or rotor in the mixer can transfer mechanical energy directly into the liquid, a large volume can be processed and high production efficiency can be achieved. We recommend that bench-scale experiments and pilot-scale production should be tried in PFD and MFD based on the lab-level experiences. For PFD, the commercial high-pressure homogenizer is recommended. For MFD, industrial rotating blade stirred tank reactors may be good choices.

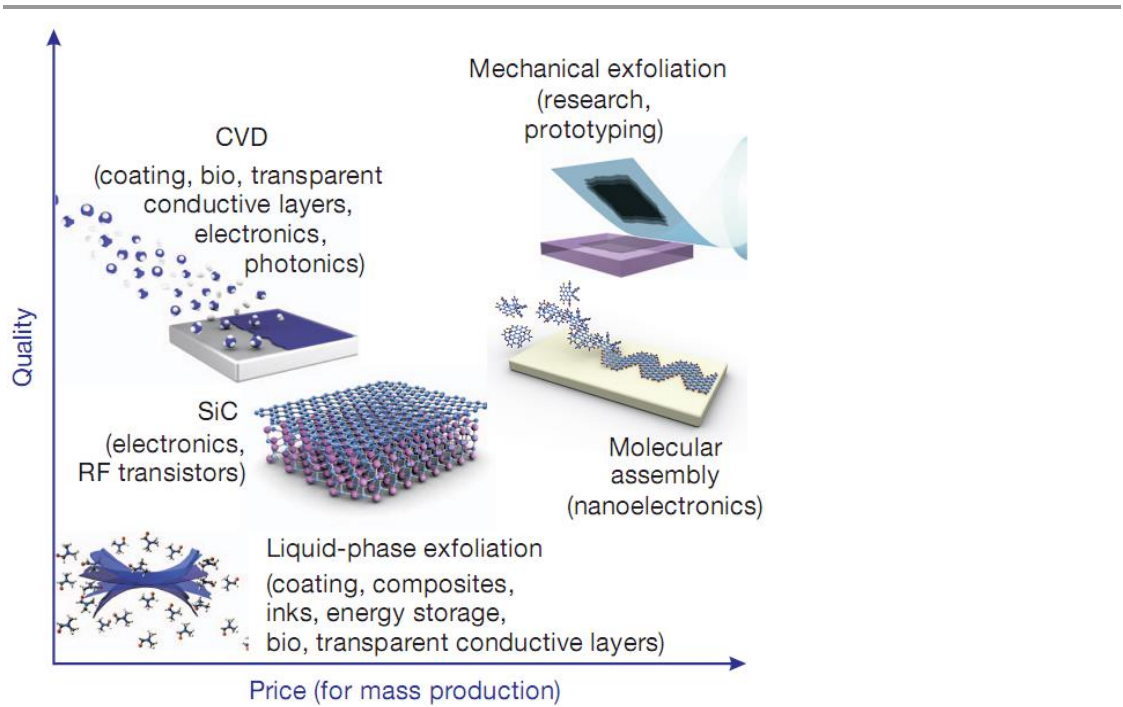

**Fig. 1.** There are several methods of mass-production of graphene, which allow a wide choice in terms of size, quality and price for any particular application. Reproduced with permission from [4]. Copyright 2012 Nature Publishing Group.

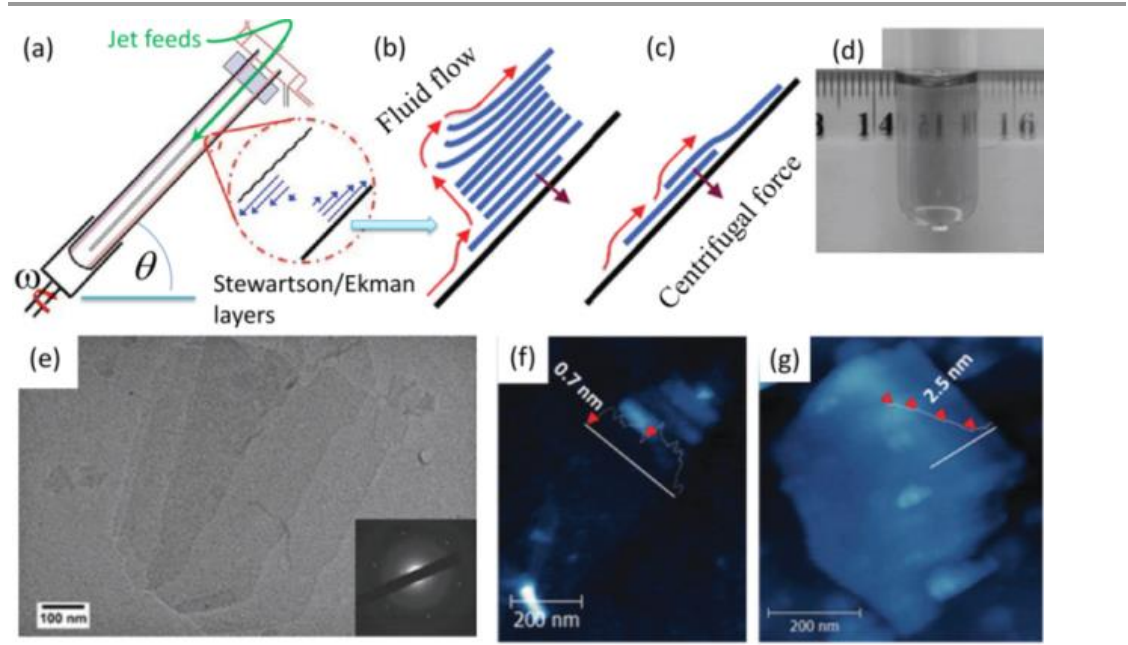

**Fig. 2.** (a) Cartoon showing the confined mode thin fluid film in the rapidly rotating tube of the VFD which has shearing force associated with the Stewartson-Ekman layers. (b) The exfoliation process with slippage and partial lift. (c) Slippage on the inner surface of the tube. (d) Photographs of the resulting colloidal suspensions of graphene (top) sheets (bottom) in NMP. (e) Partially stacked graphene for the evidence of slippage. (f) (g) AFM images of graphene. Reproduced with permission from [58] and [72]. Copyright 2012 and 2013 The Royal Society of Chemistry.

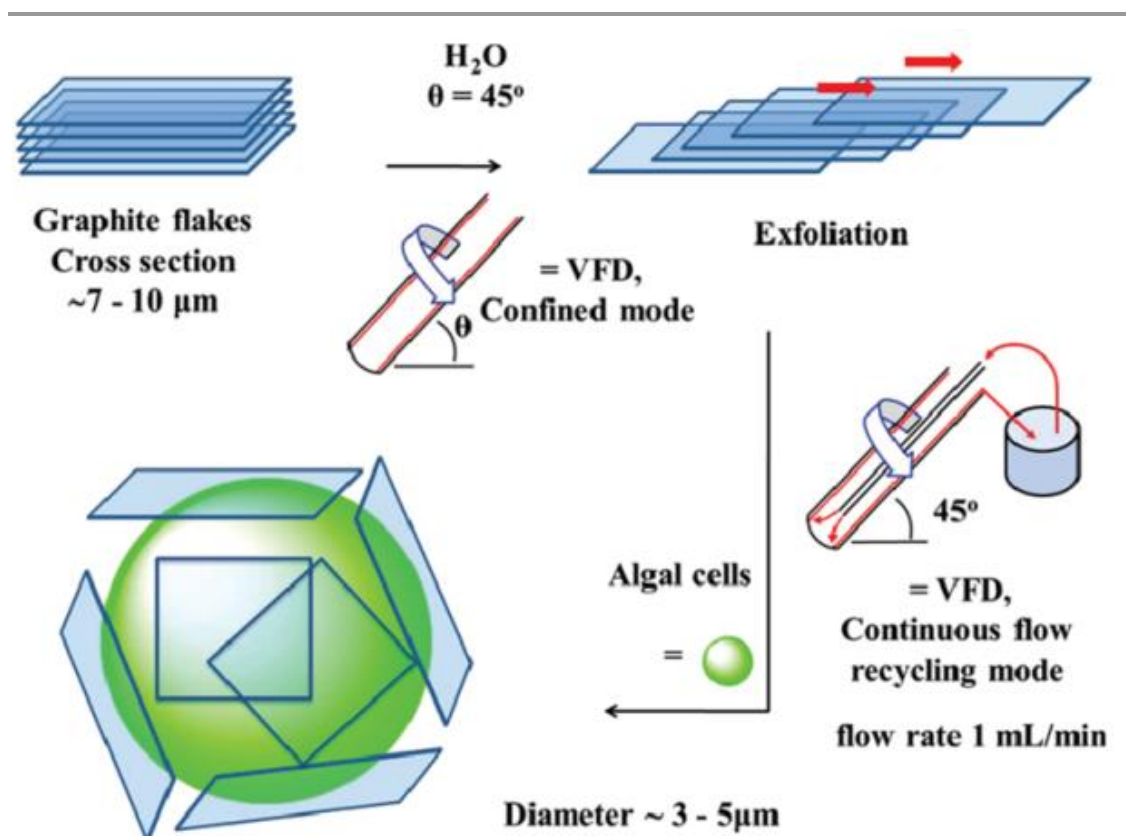

**Fig. 3.** Schematic illustration of the overall hybridization process, involving the exfoliation of graphite flakes into multi-layer graphene sheets followed by the hybridization of these sheets with algal cells using a VFD. Reproduced with permission from [72]. Copyright 2013 The Royal Society of Chemistry.

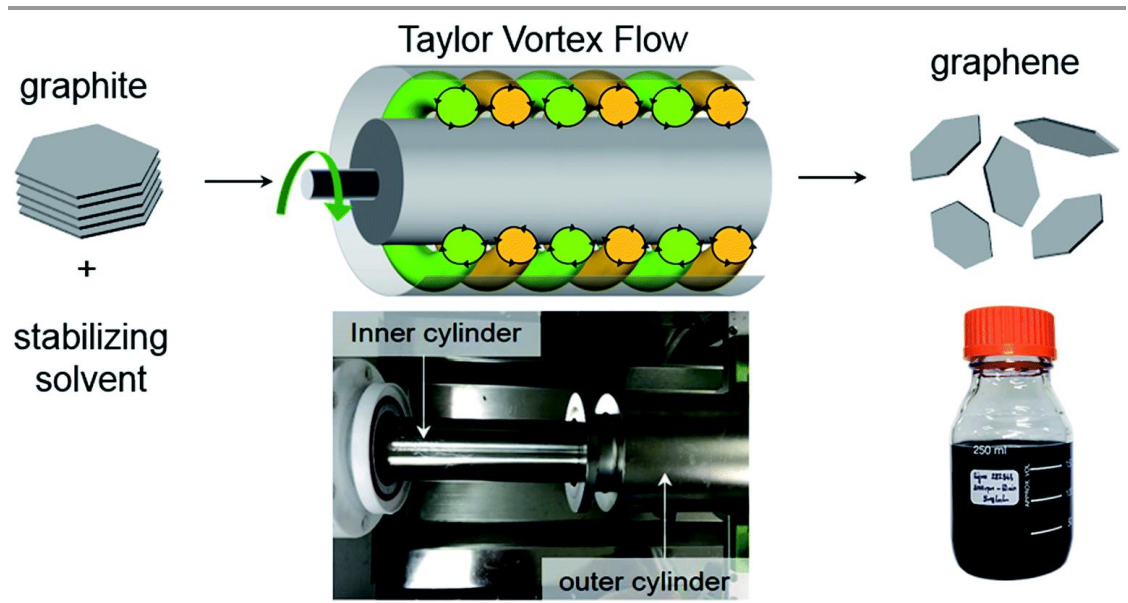

**Fig. 4.** Schematic of the exfoliation of graphite in a Taylor-Couette flow reactor. Reproduced with permission from [108]. Copyright 2016 The Royal Society of Chemistry.

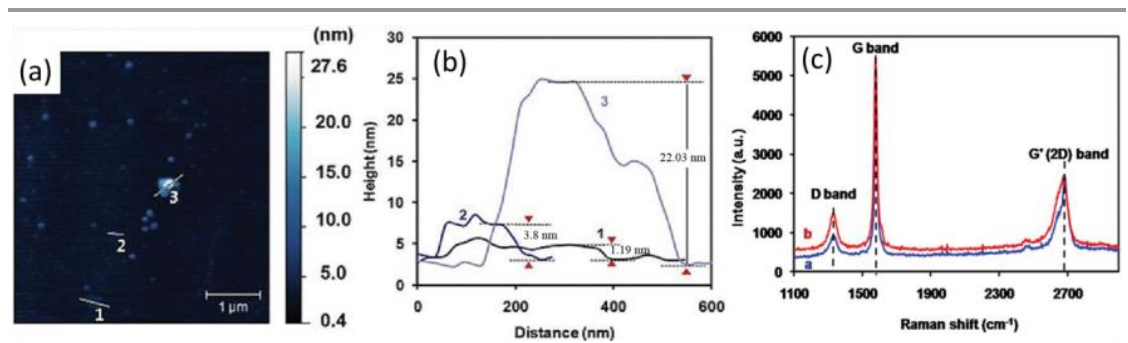

**Fig. 5.** (a) AFM image of exfoliated graphene sheets obtained in water using the VFD. (b) Height profiles of the selected area in (a). (c) Raman spectra of graphite flakes (a) before and (b) after VFD processing. Reproduced with permission from [72]. Copyright 2013 The Royal Society of Chemistry.

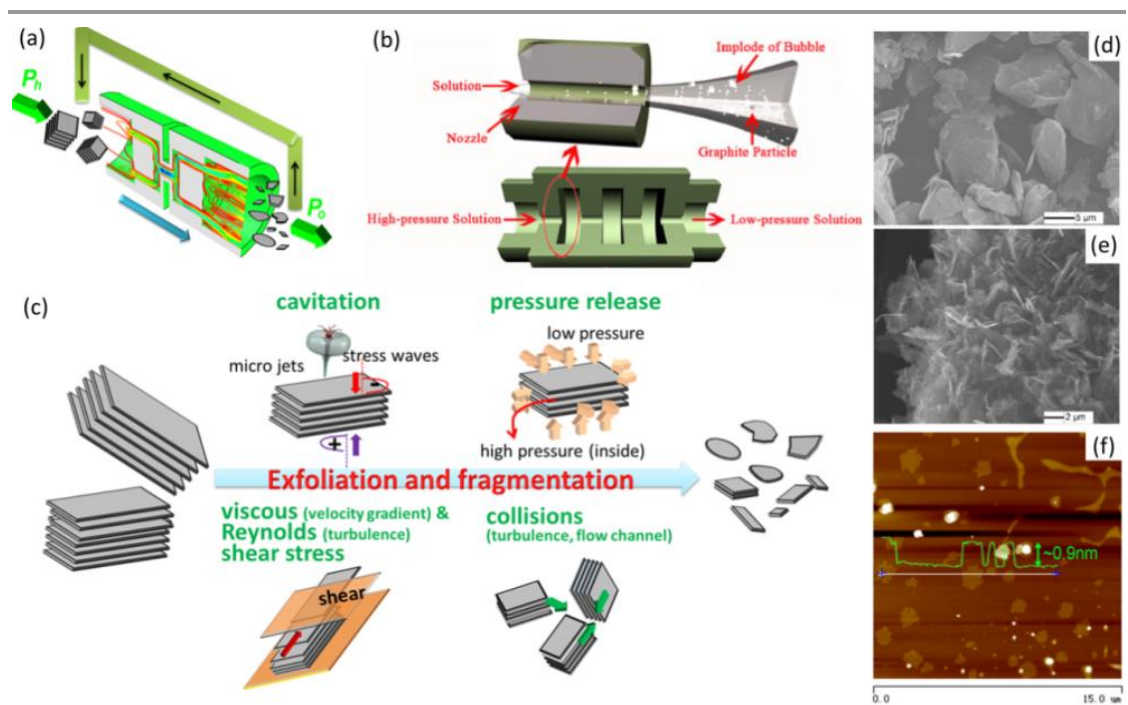

**Fig. 6.** (a) Schematic of the apparatus with one constriction channel for producing graphene. High pressure ($P_h$) is exerted by a plunger pump and $P_o$ denotes ambient pressure. (b) Schematic of the apparatus with four constriction channels. (c) Schematic of the exfoliation mechanism of the pressure driven fluid dynamics. SEM images of (d) graphite particles and (e) graphene flakes produced by the apparatus in (b). (f) AFM image of the graphene sheets prepared by the apparatus in (a). Reproduced with permission from [69] and [84]. Copyright 2013 and 2014 Springer.

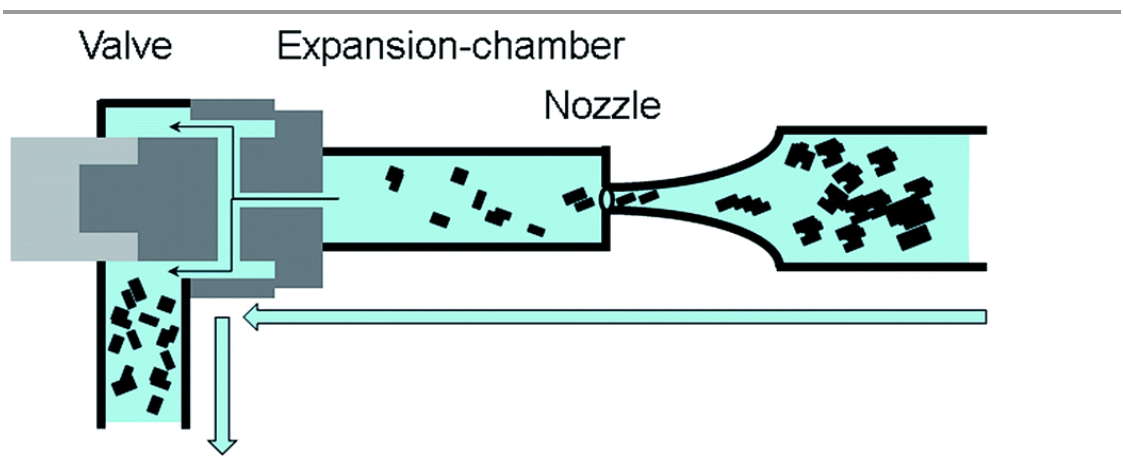

**Fig. 7.** Illustration of graphite delamination by high pressure homogenizer. The suspension is pumped through a nozzle and released into an expansion-chamber. Reproduced with permission from [109]. Copyright 2015 The Royal Society of Chemistry.

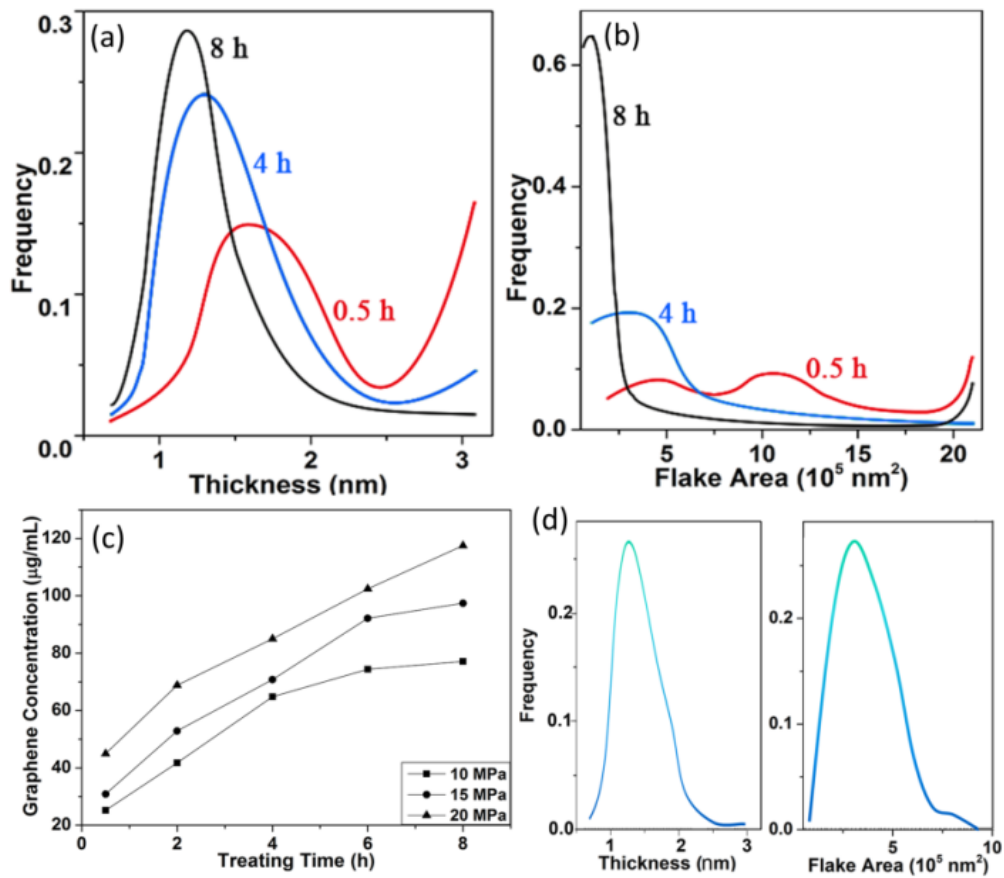

**Fig. 8.** The distribution of graphene (a) thickness and (b) flake area for different treating time of 0.5 h, 4 h, and 8 h under a pressure of 15 MPa. (c) Graphene concentration as a function of treating time and pressure. (d) Graphene thickness and flake area distribution under a pressure of 20 MPa and a treating time of 4 h. Reproduced with permission from [86]. Copyright 2015 American Scientific Publishers.

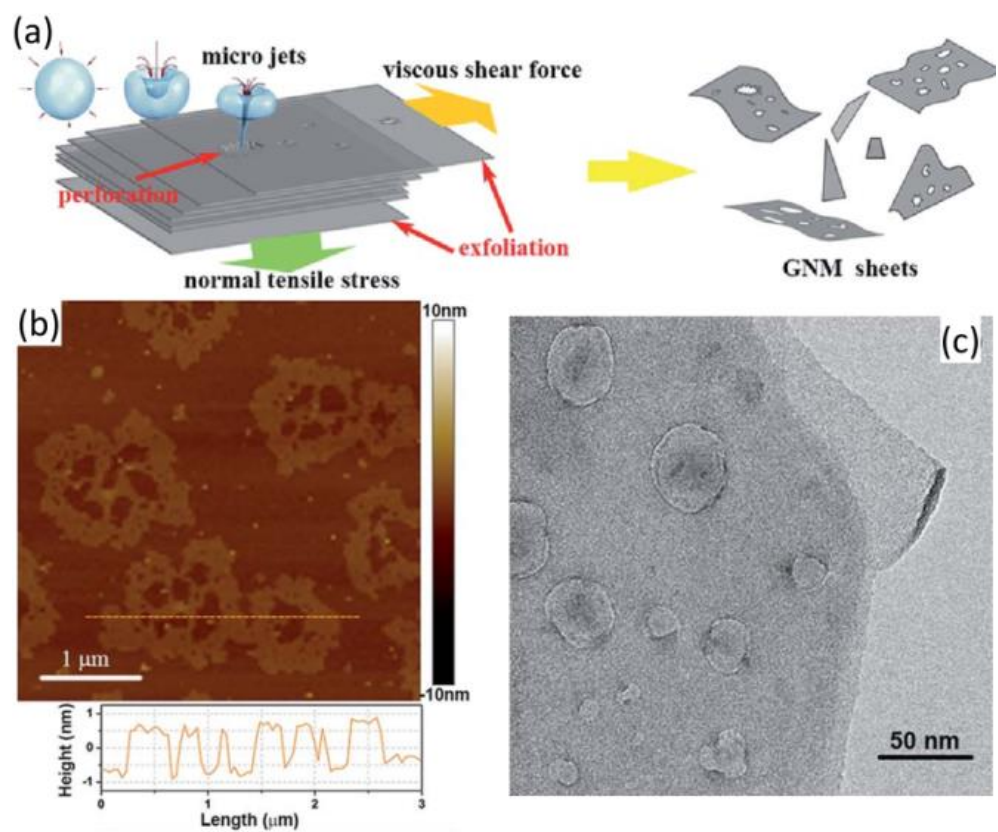

**Fig. 9.** (a) A schematic illustration for the pressure driven fluid dynamics for preparing graphene nanomesh. Typical (b) AFM and (c) TEM images of as-produced graphene nanomesh. Reproduced with permission from [78]. Copyright 2014 The Royal Society of Chemistry.

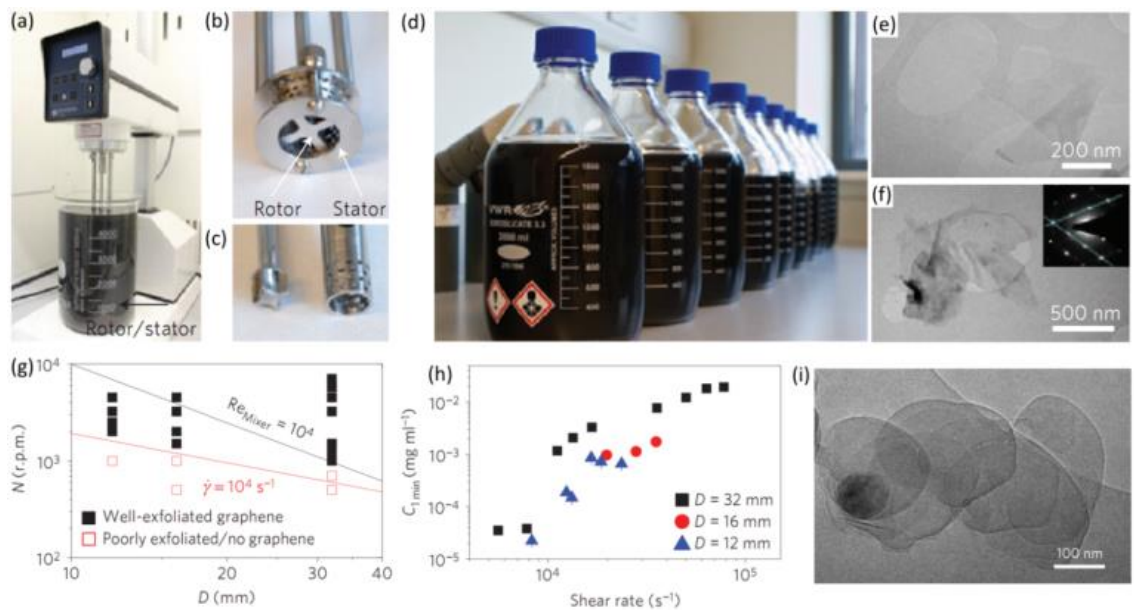

**Fig. 10.** (a) A Silverson model L5M high-shear mixer with mixing head in a 5L beaker of graphene dispersion. (b) and (c) Mixing head with rotor and stator. (d) Graphene-NMP dispersions. (e) TEM images of an individual nanosheet. (f) Multilayer graphene (bottom left) and monolayer graphene (right) as evidenced by its electron diffraction pattern (inset). (g) Phase diagram of rotor speed, $N$, versus the mixing head diameter, $D$, for dispersions showing good exfoliation. The region above the black line represents fully developed turbulence, that is, $Re_{Mixer} > 10^4$, whereas the region above the red line represents $\dot{\gamma}_{min} > 10^4$ s$^{-1}$. (h) Concentration of graphene as a function of shear rate for rotors with diameters of 32, 16 and 12 mm (mixing time 1min). All of three data sets are consistent with the same minimum shear rate. (i) TEM image of partially exfoliated BN flake, consistent with exfoliation by shear sliding. Reproduced with permission from [80]. Copyright 2014 Nature Publishing Group.

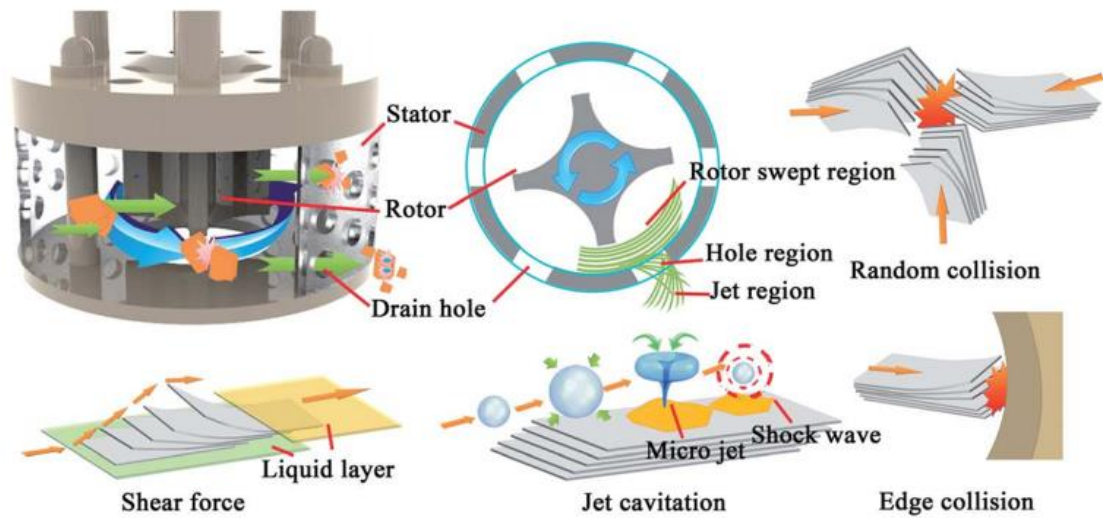

**Fig. 11.** 3D sectional drawing of the high-shear mixer, and the schematic mechanical mechanism for preparing graphene by shear force, collision, and cavitation. Reproduced with permission from [79]. Copyright 2014The Royal Society of Chemistry.

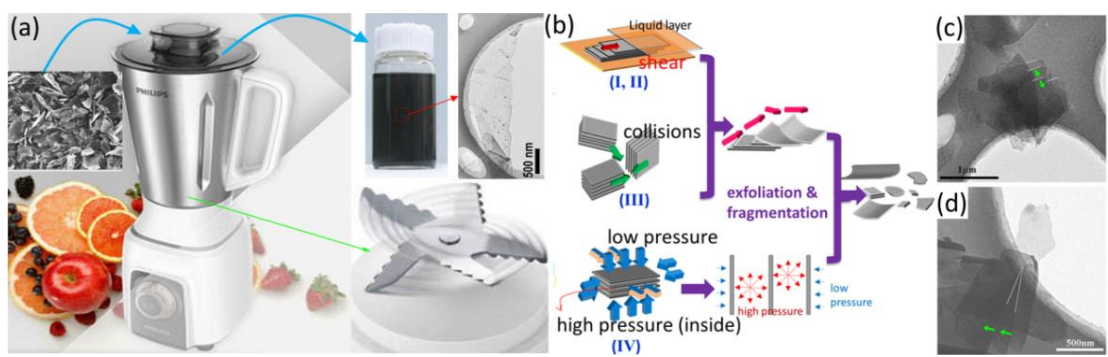

**Fig. 12** (a) The schematic of a kitchen blender for preparing graphene flakes, with DMF as the solvent. (b) Illustration for the exfoliation mechanism. Deliberately captured partially exfoliated graphene flakes with translational (c) and rotational (d) lateral exfoliation. Reproduced with permission from [83]. Copyright 2014 Elsevier.

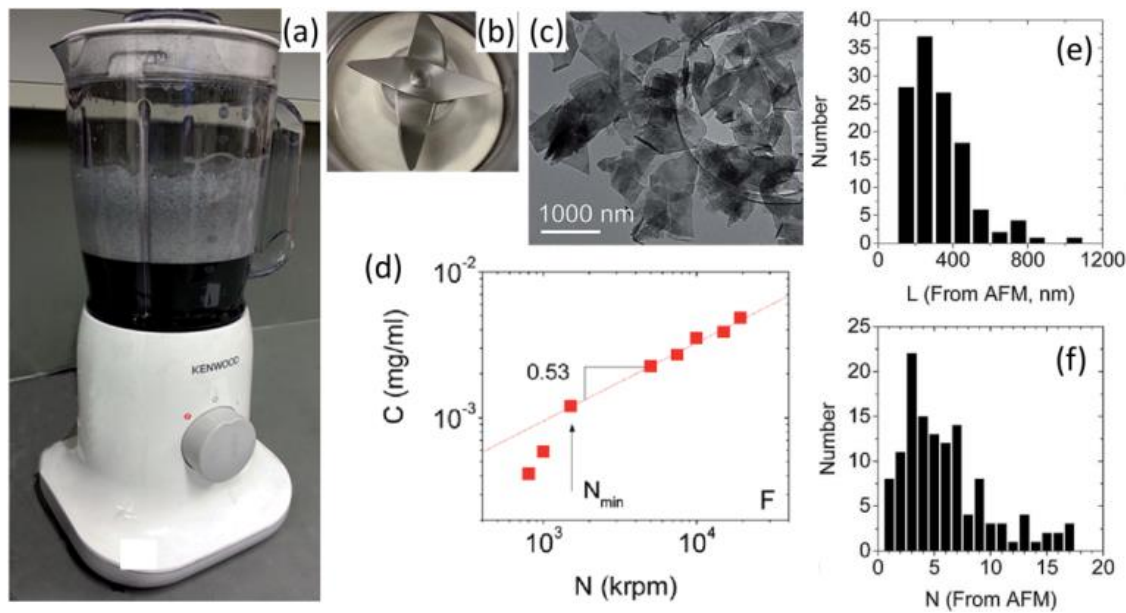

**Fig. 13** (a) The Kenwood BL370 series kitchen blender used in this work photographed during a mix. The blender is mixing graphite powder in an aqueous surfactant solution. The surfactant is the household detergent, Fairy Liquid. Note the presence of significant amounts of foam. (b) A photograph of the rotating blade supplied with this blender. (c) TEM image of graphene nanosheets. (d) Concentration of mixer dispersed graphene plotted versus blade speed N. (e) Length (L) and (f) layer number (N) distributions of flakes as measured by AFM Reproduced with permission from [81]. Copyright 2014 The Royal Society of Chemistry.

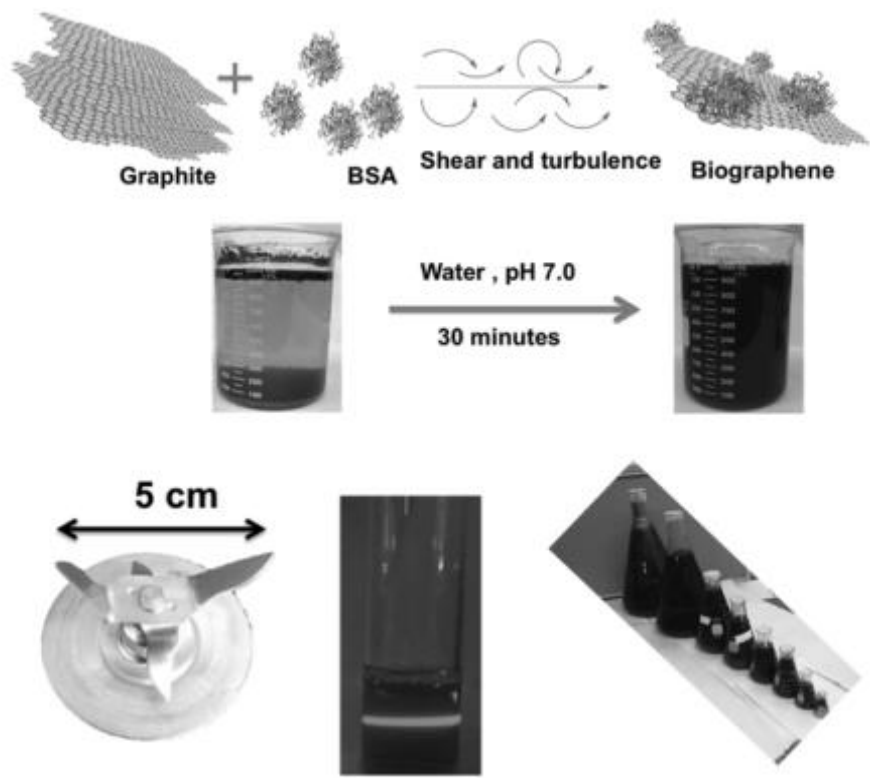

**Fig. 14.** Illustration of exfoliating graphite in aqueous solutions of proteins by a kitchen blender. 8 L solutions are processed for demonstrating the scalability of the method. Reproduced with permission from [110]. Copyright 2015 Wiley.

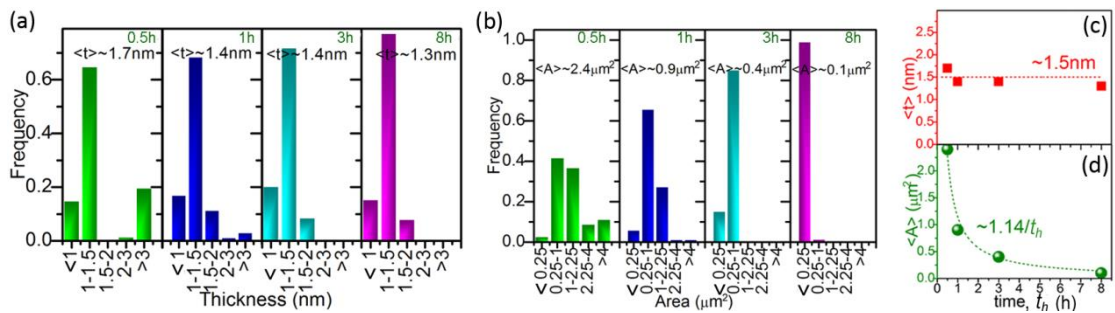

**Fig. 15.** Size distribution of graphene prepared in DMF by MFD in a kitchen blender. Statistical histogram derived from plenty graphene flakes showing the thickness (a) and area (b) distribution. Calculated (c) average thickness, <t>, and (d) average area, <A>, as a function of treating time. Reproduced with permission from [83]. Copyright 2014 Elsevier.

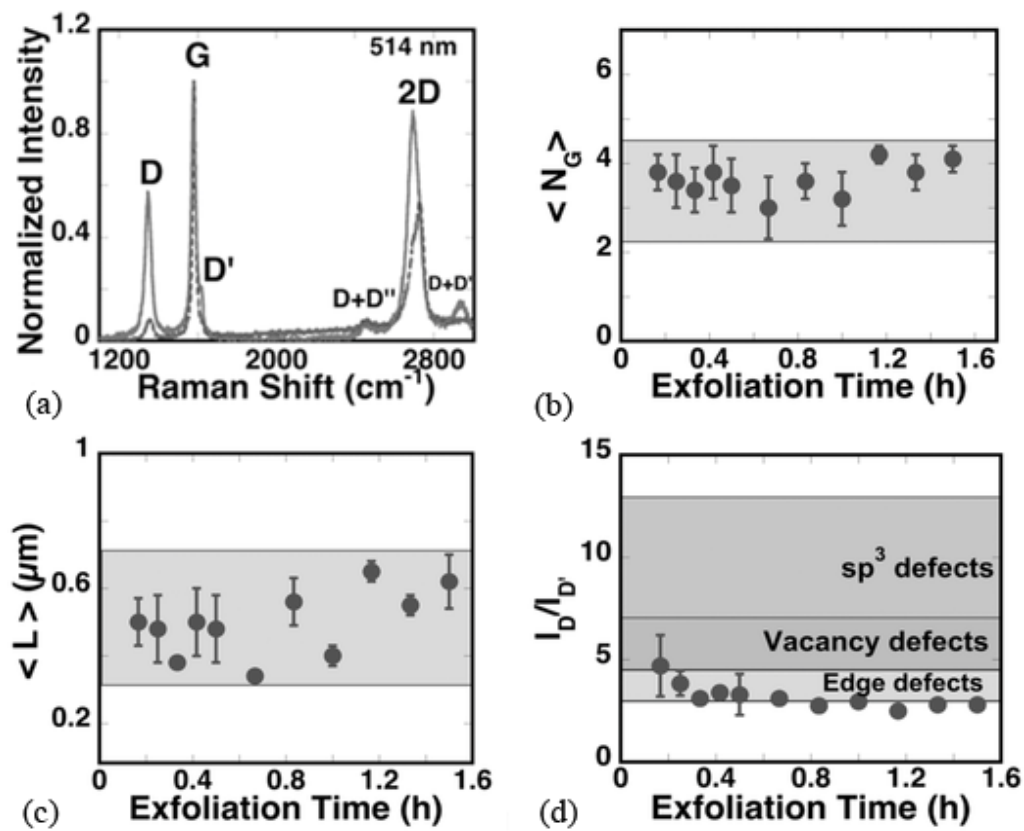

**Fig. 16**. Size and defect analysis of graphene produced in water/proteins solutions by MFD in a kitchen blender. (a) Raman spectra of graphene (solid line) and graphite (dotted line). (b) Layer number and (c) lateral size as a function of exfoliation time. (d) Statistic analysis of Raman results of graphene showing the defect information. Reproduced with permission from [110] Copyright 2015 Wiley.

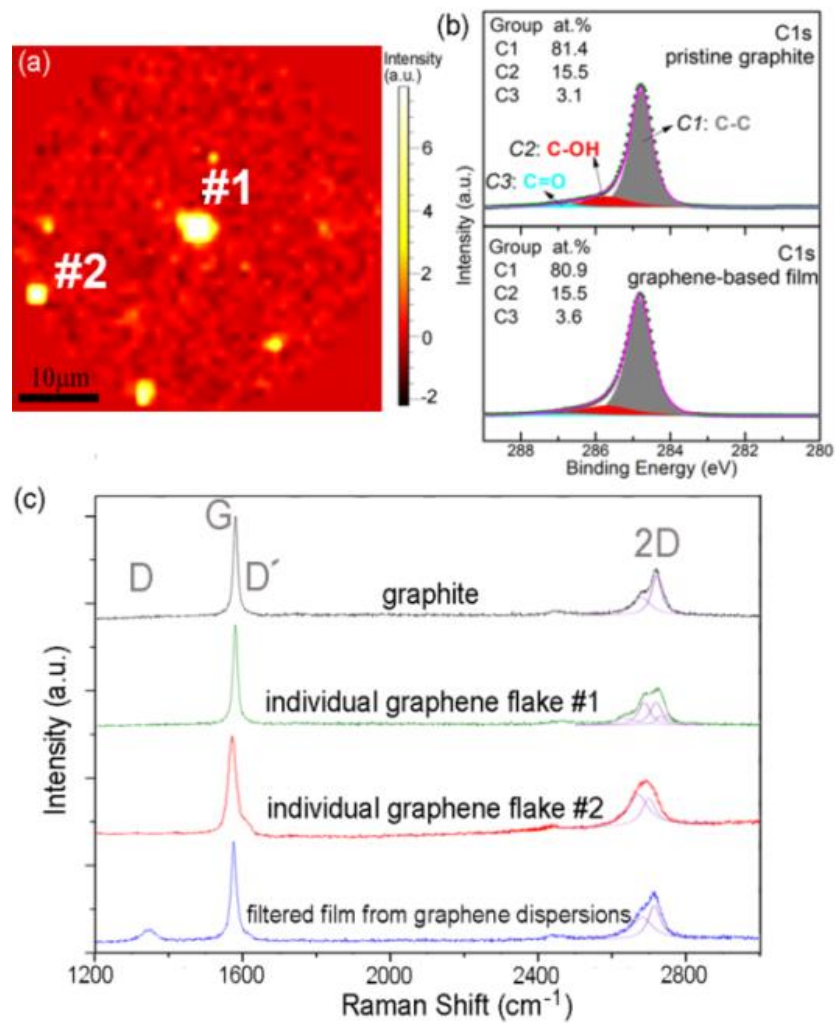

**Fig. 17.** Defect analysis of graphene produced in DMF by MFD in a kitchen blender. (a) A Raman mapping image. The Raman map plots the intensity integral of the spectra between 2600 and 2800 cm$^{-1}$. The excitation wavelength is 532 nm. (b) Carbon 1s core-level XPS spectra of the pristine graphite and graphene-based film. (c) Raman spectra for bulk graphite, individual flake #1, individual flake #2, and the filtered film. Reproduced with permission from [83]. Copyright 2014 Elsevier.

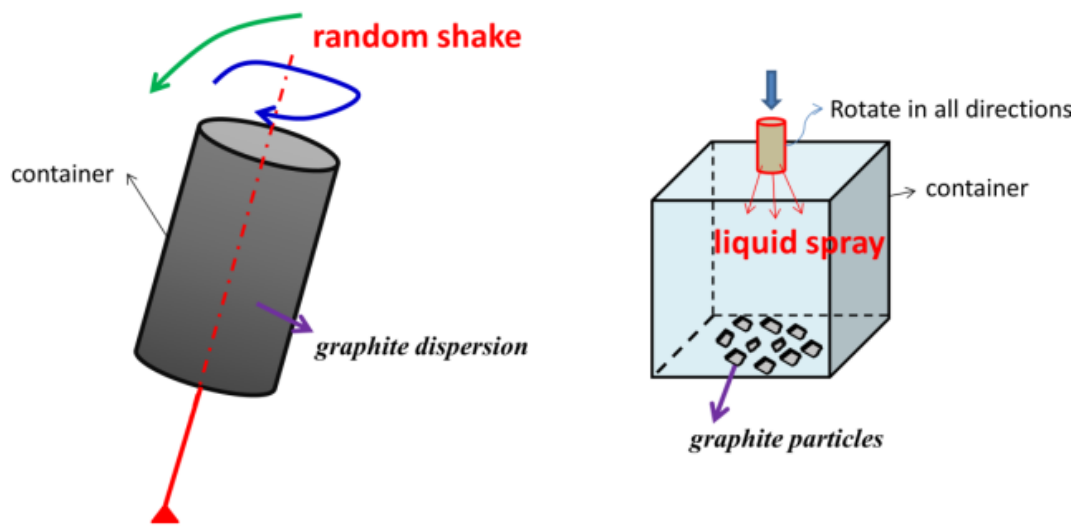

**Fig. 18.** Schematics of another two possible or presumptive routes for generating fluid dynamics for graphene production: random shake and liquid spray.